\documentclass[runningheads]{cl2emult}

\usepackage{graphicx} 
\usepackage{subeqnar} 
\usepackage{cropmark} 
\usepackage{physbb}   

\begin{document}
\title*{Soft and hard pomerons}
\toctitle{Soft and hard pomerons}
%
%
\titlerunning{Soft and hard pomerons}
%
\author{P V Landshoff\inst{1}
\authorrunning{P V Landshoff}
%
%
\institute{DAMTP, University of Cambridge, Cambridge CB3 9EW, England}
}
\maketitle              

\begin{abstract}
Regge theory provides an excellent description of small-$x$ structure-function
data from $Q^2=0$ up to the highest available values. The large-$Q^2$ data
should also be described by perturbative QCD: the two descriptions must 
agree in the region where they overlap. However, at present there is
a serious lack in our understanding of how to apply perturbative QCD at
small $x$. The usual lowest-order or next-to-lowest order expansion is
not valid, at least not until $Q^2$ becomes much larger than is usually
assumed; a resummation is necessary, but as yet we do not know how to
do this resummation.
\end{abstract}
\def\e{\epsilon}
\def\half{\textstyle{{1\over 2}}}
\section{Introduction}
Perturbative QCD has become a well-established description of hard
processes. But it is not complete:  it must be supplemented
with other descriptions. 

Regge theory is an example of another description. It was extensively
developed some 40 years ago\cite{collins} and is based on our knowledge
of analyticity properties of scattering amplitudes. It relates the
high-energy behaviour of scattering amplitudes to exchanges of known
particles. In order to describe the data, it also introduces 
extra terms that, at
least so far, are not related to exchanges of known particles. These
extra terms are called pomeron-exchange terms, after the Russian
physicist Isaac Pomeranchuk. The notion of the pomeron is a very old one.
We do not know whether it too describes the exchange of particles,
but if it does there is general agreement that these
are likely to be glueballs. 

Regge theory gives an excellent description not only of soft hadronic
proceses\cite{collins}\cite{diffdis}\cite{sigtot},
but also of the behaviour of the structure function $F_2(x, Q^2)$ at
small $x$, all the way from real-photon-induced events
($Q^2=0$) up to very deeply inelastic ones (large $Q^2$). 
The structure function describes an example of a semihard process.
Another example of a semihard process is the quasi-elastic reaction
$\gamma ~p \to J/\psi ~p$.
For such semihard processes, the data reveal\cite{twopom}
that there exists a second
pomeron, the ``hard'' pomeron, in addition to the original ``soft''
pomeron which enters into purely hadronic reactions. The soft pomeron
is certainly nonperturbative in origin, though it is possible that
the hard pomeron is associated with the perturbative BFKL equation\cite{bfkl}.
However, there are many serious problems\cite{cl}\cite{fl}
with the BFKL equation, and so this
is not sure.

Certainly, one would hope that at small $x$ and large $Q^2$ Regge theory
can be made to agree with perturbative QCD, in particular with 
DGLAP evolution\cite{esw}. However, there are problems with the
DGLAP equation too at small~$x$. The equation
involves a kernel or splitting function
$P(z,\alpha_S(Q^2))$. If one expands this in powers of $\alpha_S$, each
term is singular at $z=0$. However\cite{cudell} it is rather sure, from general
considerations of the known analyticity in $Q^2$ of the structure function,
that $P(z,\alpha_S(Q^2))$ is not singular at $z=0$. The singularities in
the terms of the expansion are a signal that the expansion is illegal
near $z=0$, and a resummation is needed to remove them. At present, we do
not know exactly how to do this. It is possible that the presence or absence
of the singularity is immaterial if $Q^2$ is large enough\cite{ballforte},
but for most applications in the literature\cite{grv} it is likely to be a real
problem.

\section{Regge theory}
Through the optical theorem, the structure function $F_1(x, Q^2)$
is the imaginary part of the virtual Compton amplitude
$T_1(\nu, t, Q^2)$ evaluated at zero momentum transfer, $t=0$. Here
$2\nu =2p.q=Q^2/x$. Regge theory begins\cite{collins}
by considering the crossed-channel
process $\gamma ^*\gamma ^*\to p\bar p$, for which $\sqrt t$ is the centre-of-mass
energy. It first makes a partial-wave series expansion in this channel, in terms
of partial-wave amplitudes $a_{\ell}(t,Q^2)$ and Legendre polynomials
$P_{\ell}(\cos\theta _t)$, where $\theta _t$ is the crossed-channel
scattering angle.  This expansion has a definite, but limited, region of
convergence in the space of the three variables $\cos\theta _t,t,Q^2$ or,
equivalently, $\nu, t,Q^2$. The partial-wave amplitude $a_{\ell}(t,Q^2)$
is defined initially for physical values of the angular momentum,
$\ell =0,1,2,\dots$. By a well-defined procedure, its definition is extended
to all values of $\ell$, both real and complex, and the partial wave series
is converted to an integral: 
\begin{equation}
T_1(\nu,t,Q^2)={1\over 2i}\int_Cd\ell{(2\ell +1)P_{\ell}(-\cos\theta _t)
\over\sin \pi\ell }a({\ell},t,Q^2)
\label{watson}
\end{equation}
\begin{figure}
\begin{center}
\includegraphics[width=.8\textwidth]{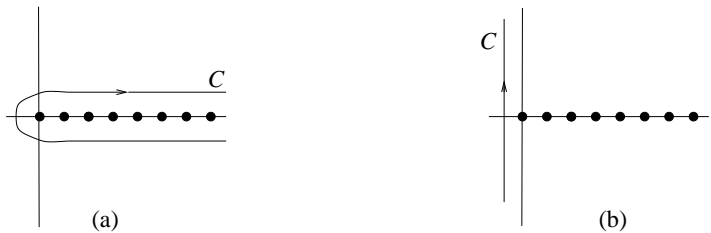}
\end{center}
\caption[]{Contour $C$ in the complex $\ell$ plane for the
integral (\ref{watson})}
\label{eps1}
\end{figure}
The contour $C$ is initially that of figure 1a,
wrapped arround the positive real-$\ell$ axis on which are located the zeros
of the denominator $\sin\pi\ell$. But the properties of $a({\ell},t,Q^2)$
allow one to use Cauchy's theorem to
distort it to become parallel to the imaginary-$\ell$ axis,
as in figure 1b. It turns out that this extends the 
region of convergence beyond that of the original
series, and the integral can be continued analytically to where we need it to
calculate the structure function $F_1(x, Q^2)$:
\begin{equation}
t=0~~~~~~~~~~~~2\nu > Q^2\geq 0
\end{equation}
As we continue analytically into this region, the singularities of 
$a({\ell},t,Q^2)$ will move around in the complex $\ell$-plane, and one of them
may try to cross the contour $C$. We must distort the contour again so
as to avoid this happening. Depending on whether the singularity is
a branch point or a pole, we then have either figure 2a or 2b.
\begin{figure}
\begin{center}
\includegraphics[width=.7\textwidth]{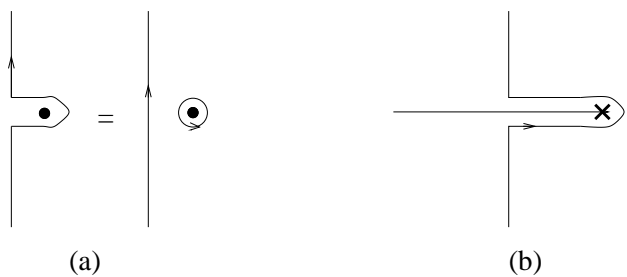}
\end{center}
\caption[]{Distortion of the contour $C$ caused by (a) a pole, or (b) a
branch point, of $a({\ell},t,Q^2)$ trying to cross it}
\label{eps2}
\end{figure}

{}From what is known about the analyticity structure, it is rather sure
that the positions of the singularities of $a({\ell},t,Q^2)$ in the complex
$\ell$ plane do not depend on $Q^2$, only on $t$.
In the case
of a pole
\begin{equation}
a({\ell},t,Q^2) \sim {\beta(Q^2,t)\over \ell-\alpha (t)}
\label{pole1}
\end{equation}
the integration around it yields a contribution
\begin{equation}
{\pi \beta(Q^2,t)\, P_{\alpha (t)}(-\cos\theta _t)\over\sin\pi\alpha(t)}
\label{pole2}
\end{equation}
to $T_1(\nu,t,Q^2)$. Now when $t=0$ the analytic continuation of
$\cos\theta _t$  is 
\begin{equation}
\cos\theta _t={\nu\over iQm_p}
\label{thetat}
\end{equation}
and when this is large the analytic continuation of the Legendre polynomial
has the simple behaviour
\begin{equation}
{\pi P_{\ell}(-\cos\theta _t)\over\sin\pi\ell}
\sim -{\Gamma(-\ell)\Gamma(\ell+\half)\over\sqrt\pi}(-2\cos\theta _t)^{\ell}
~~~~~~~~~\hbox{(Re }\ell>\half\hbox{)}
\label{legendre}
\end{equation}
So, with (\ref{thetat}), we see that
the integral (\ref{watson}) becomes just a Mellin transform, 
and the ``Regge trajectory'' $\alpha(t)$ contributes at $t=0$
\begin{equation}
b_1(Q^2)\nu^{\alpha (0)}
\end{equation}
to the large-$\nu$ behaviour of $T_1(\nu,0,Q^2)$. Here, $b_1(Q^2)$ is a constant
multiple of $\beta(Q^2,0)$. In the case of
$T_2(\nu,0,Q^2)$ its definition includes a kinematic factor which reduces the
power of $\nu$ by one unit, so
since $\nu= Q^2/2x$ this gives
\begin{equation}
F_2(x,Q^2)\sim f(Q^2)x^{-\e}
\label{f2}
\end{equation}
with
\begin{equation}
\e=\alpha(0)-1
\end{equation}
Regge theory gives no information about the function $f(Q^2)$, other
than that it is an analytic function with singularities whose locations
are known\cite{elop}. The power $(1-\alpha (0))$ is independent of $Q^2$.

\section{Fit to data}
In the case where the singularity that crosses the contour $C$ is a branch
point instead of a pole,  dragging a branch cut with it as shown in
figure 2b, the simple power of $x$ in (\ref{f2}) is replaced with
something more complicated. One knows, from unitarity\cite{collins}, 
that if there are poles in
the complex $\ell$ plane, there must certainly also be branch points.

On the principle that it is usually the best strategy to try the simplest
possible assumption first, Donnachie and I
tested the hypothesis that at $t=0$ the contribution
from branch points is much weaker than from poles. Not
everybody agrees with this strategy\cite{capella}, but we applied this 
several years ago\cite{sigtot}
to purely hadronic total cross-sections and found that
they are all described well by a sum of just two powers $\nu^{\e _1}$
and $\nu^{\e _2}$ The two powers are
\begin{equation}
\matrix{\e _1=0.08&\hbox{(``soft pomeron'' exchange)}\cr
\e _2=-0.45&\hbox{($\rho,\omega, f,a$ exchange)}\cr}
\label{powers}
\end{equation}

More recently\cite{twopom}, we tested the hypothesis
that also the contribution to the structure function $F_2$ at small $x$
from branch points is much weaker than from poles. We made a fit of the
form
\begin{equation}
F_2(x,Q^2)\sim\sum _{i=0}^2f_i(Q^2)x^{-\e _i}
\label{f2sum}
\end{equation}
We fixed the values of the powers $\e _1$ and $\e _2$ 
to be the same as in (\ref{powers}), and left $\e _0$ as a free parameter
to be determined from the small-$x$ structure-function data.
Our fitting procedure had three stages. First, we arrived at a
provisional value for $\e _0$ by using data only in the region
\begin{equation}
x<0.07~~~~~~~~~~~~~~~~~~~0\leq Q^2<10~\hbox{GeV}^2
\end{equation}
which gave us 
\begin{equation}
\e _0\approx 0.4
\end{equation}
Next, with this value for $\e _0$, we fitted the data to a sum 
(\ref{f2sum}) of three powers of $x$ for each values of $Q^2$ for
which there exist data, still restricting $x$ to less than 0.07$\,.$
This gave us the plots shown in figure 3 of the hard and soft pomeron
coefficient functions $f_0(Q^2)$ and $f_1(Q^2)$ as functions of $Q^2$.
The data do not constrain the ($f,a$)-exchange coefficient function
$f_3(Q^2)$ at all well, so we retained the simple form we chose for it 
in the first stage of our fitting procedure. 
\begin{figure}
\begin{center}
\includegraphics[width=.8\textwidth]{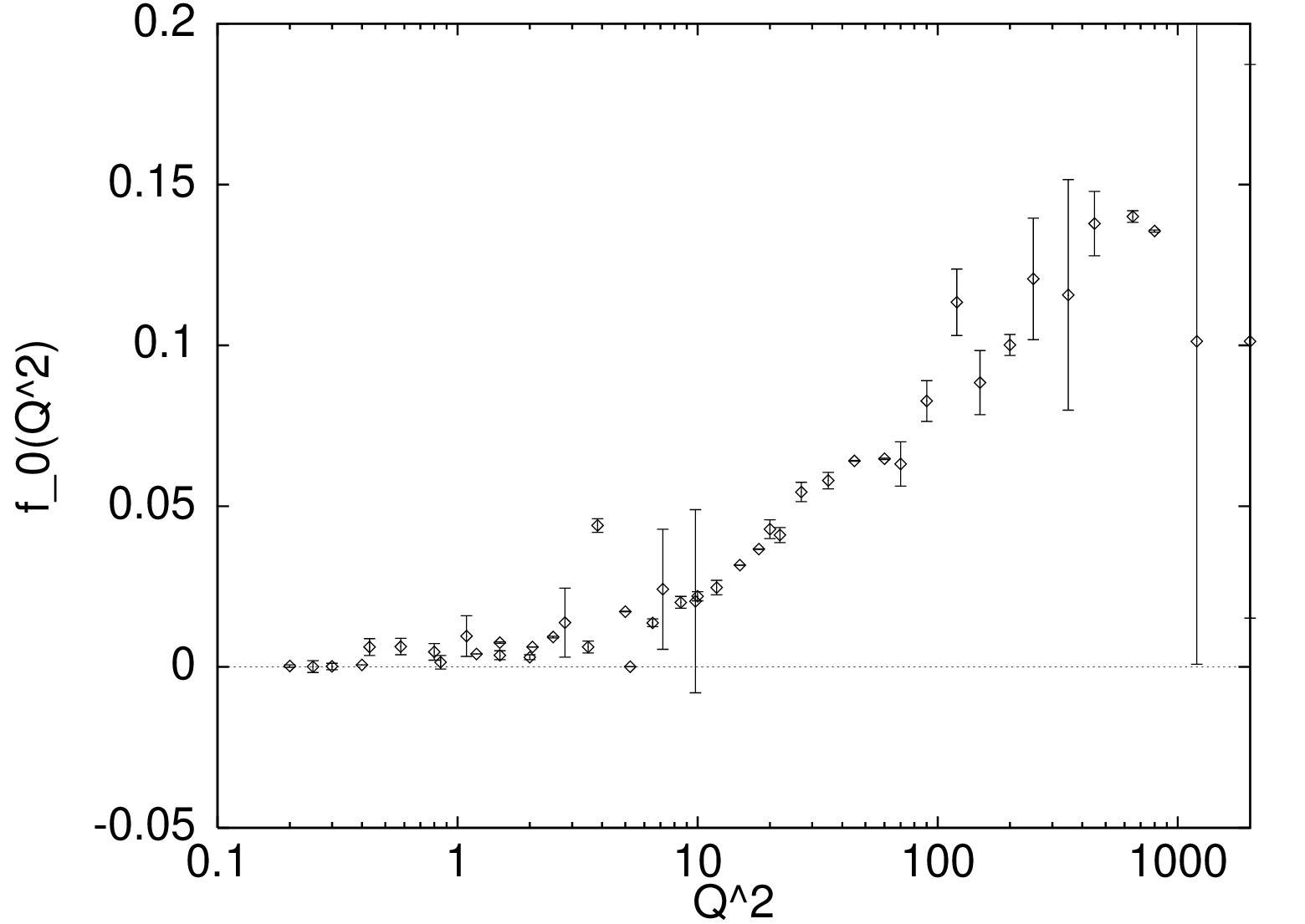}
\vskip 6mm
\includegraphics[width=.8\textwidth]{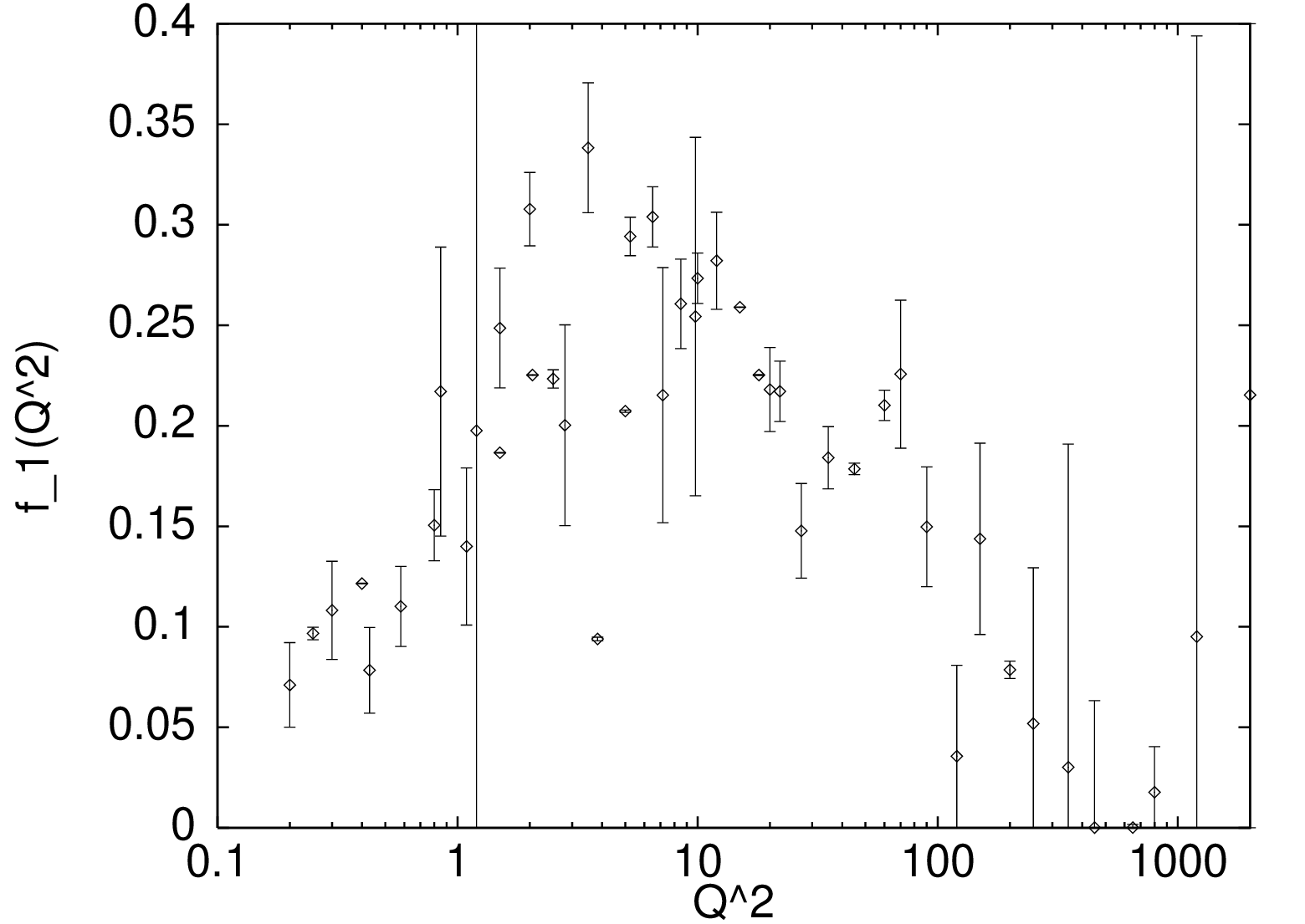}
\end{center}
\caption[]{The hard and soft pomeron coefficient functions $f_0(Q^2)$
and $f_1(Q^2)$ extracted from the data}
\label{eps3}
\end{figure}

We know from gauge invariance that $F_2$ must
vanish linearly in $Q^2$ as $Q^2\to 0$ at fixed $\nu$. 
We see that the hard-pomeron
coefficient function $f_0(Q^2)$ recovers from this rather slowly as
$Q^2$ increases, until it begins at $Q^2\approx 10$ GeV$^2$ to
rise quite rapidly. On the other hand, the soft--pomeron
coefficient function $f_1(Q^2)$ rises rapidly away from $Q^2=0$,
until it peaks at between 5 and 10 GeV$^2$. It then falls again. This
was a surprise to us: the soft-pomeron contribution to the small-$x$
structure function apparently is higher twist. 

The third part of our fitting procedure chooses functions that have the
general shape of the plots in figure 3, with a number of parameters. It
still retains the same form for $f_2 (Q^2)$, which matters significantly
only for the real-photon data at low energy --- our fit includes these
for $\sqrt s > 6$ GeV, and including the $f_2$ term is essential to
fit them. So we use (\ref{f2sum}) with
\begin{eqnarray}
f_0(Q^2)&=A_0\left ({Q^2\over Q^2+Q_0^2}\right )^{1+\e _0}
             \left (1+{Q^2\over Q_0^2}\right )^{\half\e _0}\\
f_1(Q^2)&=A_1\left ({Q^2\over Q^2+Q_1^2}\right )^{1+\e _1}
             \left (1+\sqrt{Q^2\over Q_S^2}\right )^{-1}\\
f_2(Q^2)&=A_2\left ({Q^2\over Q^2+Q_2^2}\right )^{1+\e _2}
\label{coeff}
\end{eqnarray}
The hard-pomeron power is now again a free parameter, together with
the three coefficients $A_i$, the mass scales $Q_i^2$, and $Q_S^2$.
With these 8 parameters, we obtain a $\chi ^2$ per data point of 1.0
for 595 data points. These data points have $x<0.07$ and range from
$Q^2=0$ to 2000 GeV$^2$. Sample plots are shown in figure 4.
\begin{figure}
\begin{center}
\includegraphics[width=.68\textwidth]{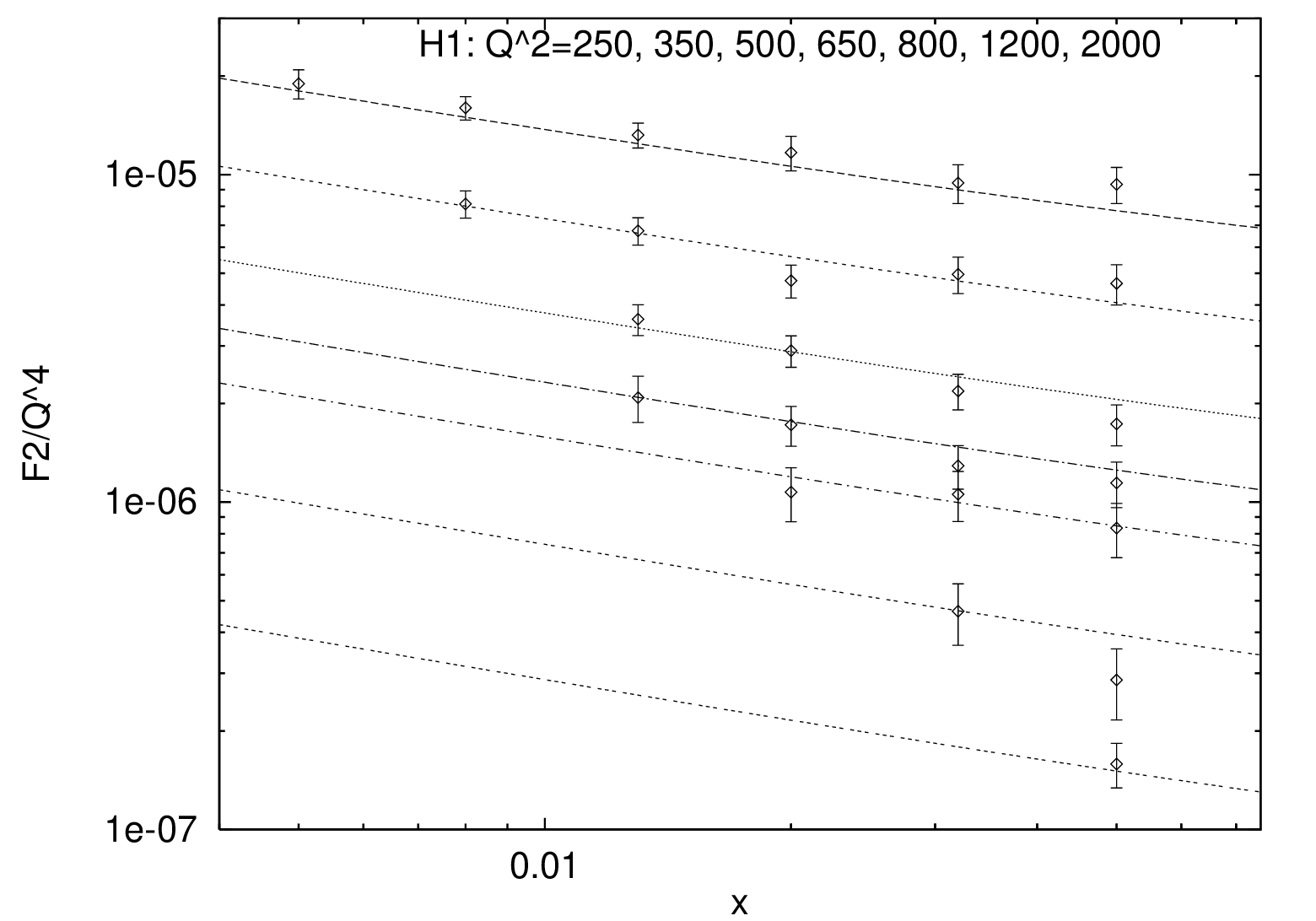}
\vskip 3mm
\includegraphics[width=.68\textwidth]{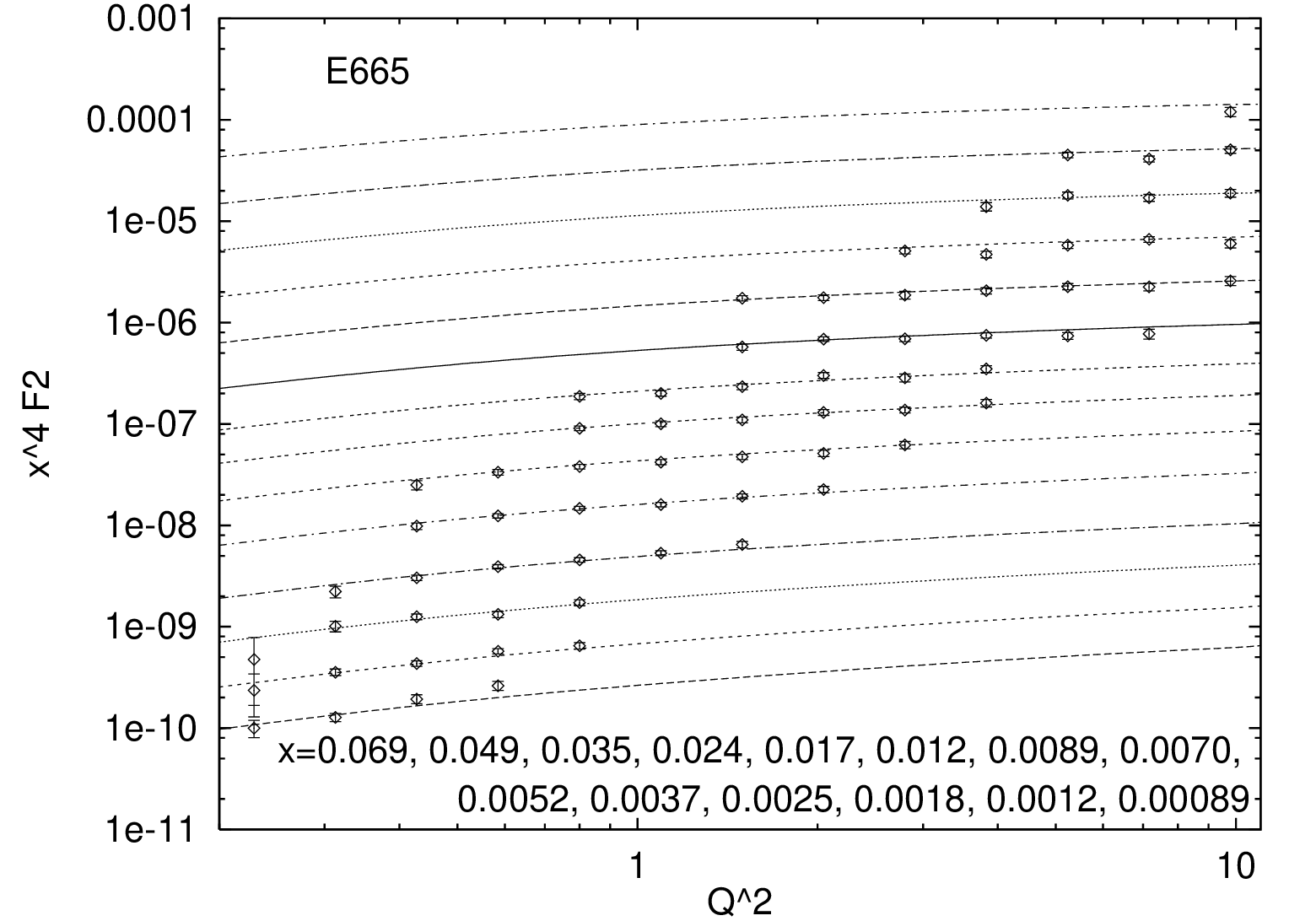}
\vskip 3mm
\includegraphics[width=.68\textwidth]{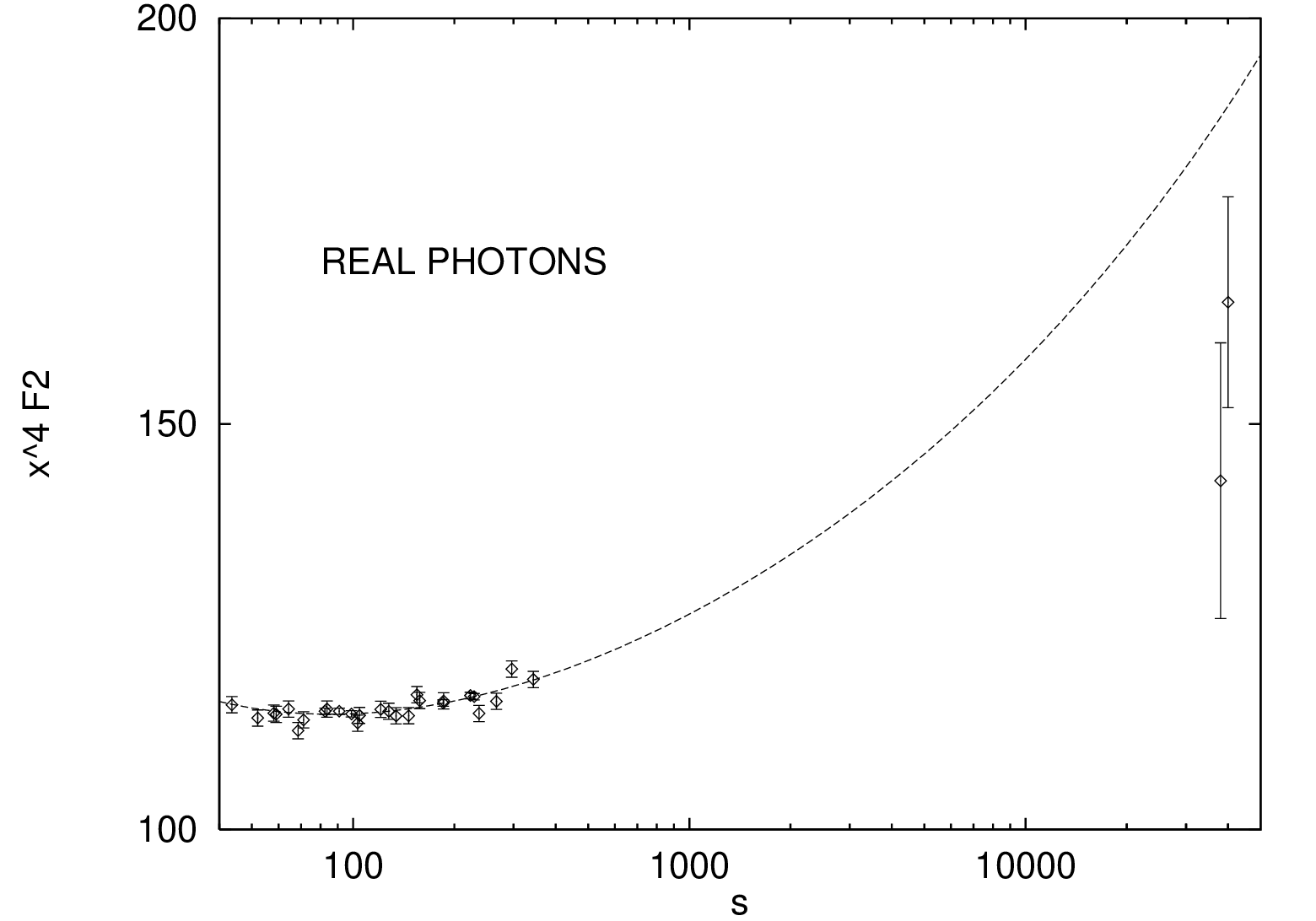}
\end{center}
\caption[]{The Regge fit compared with some of the data, at the largest
available $Q^2$, small $Q^2$, and $Q^2=0$}
\label{eps4}
\end{figure}

I want to make a number of comments on these fits:
\begin{itemize}
\item The choice (\ref{coeff}) for the analytic form of the coefficient
functions is an economical set that describes the data well, but there
are other choices that agree also with the data extracted in figure 3.
Using different choices results in different values for the hard-pomeron
power $\e_0$, but they are all within 10\% of 0.4.
\item The choice (\ref{coeff}) makes $f_0(Q^2)\sim Q^{\e_0}$ at large
$Q^2$, which corresponds to $\beta (Q^2,0)$ in (\ref{pole1}) becoming
constant for large $Q^2$. There is no general theory that explains this,
though it has been predicted\cite{ermolaev} from the BFKL equation.
Parametrisations of $f_0(Q^2)$ that behave logarithmically at large $Q^2$
can also fit the data satisfactorily.
\item The choice (\ref{coeff}) makes $f_1(Q^2)\sim 1/Q$ at large
$Q^2$. Fits that make it instead behave as $1/Q^2$ are also
acceptable, though less good. A $1/Q$ behaviour has been predicted\cite{levin}
from a combination of the BFKL equation  and infrared renormalons,
though this is controversial\cite{sotiropoulos}. Fits in which $f_1(Q^2)$
does not go to zero at high $Q^2$ do not work: it really does seem that
the soft-pomeron contribution to $F_2(x,Q^2)$ is higher twist. At, say,
$Q^2=5$ GeV$^2$ soft-pomeron exchange dominates in  $F_2(x,Q^2)$ 
until $x$ is less than about $10^{-3}$. The consequence of this for
conventional structure-function fits\cite{grv} needs discussion.
\item Without the contribution from the hard pomeron, the fit to the
real-photon data agrees well\cite{sigtot} with the measured HERA points.
However, it is seen in figure 4 that including the hard pomeron makes
the fit pass significantly above these. 
\item Figure 5 shows a fit to the preliminary ZEUS data for the charm structure
function. It shows that, to a good approximation, these data may be described
well by hard-pomeron exchange alone. The fit is actually a single-parameter
fit; it includes the constraint that at high $Q^2$ the hard pomeron is
flavour blind.
\item Figure 6 shows that the data for the semihard process 
$\gamma\,p\to J/\psi\,p$ are well described by a mixture of soft and
hard pomeron exchange\cite{twopom}. The figure is for $Q^2=0$; the HERA
data indicate that, as $Q^2$ increases, the hard-pomeron component becomes
relatively more important. One might guess that the data for 
$\gamma\,p\to \rho\,p$ may be described similarly, though at $Q^2=0$
the soft pomeron dominates.
\end{itemize}
\begin{figure}
\begin{center}
\includegraphics[width=.6\textwidth]{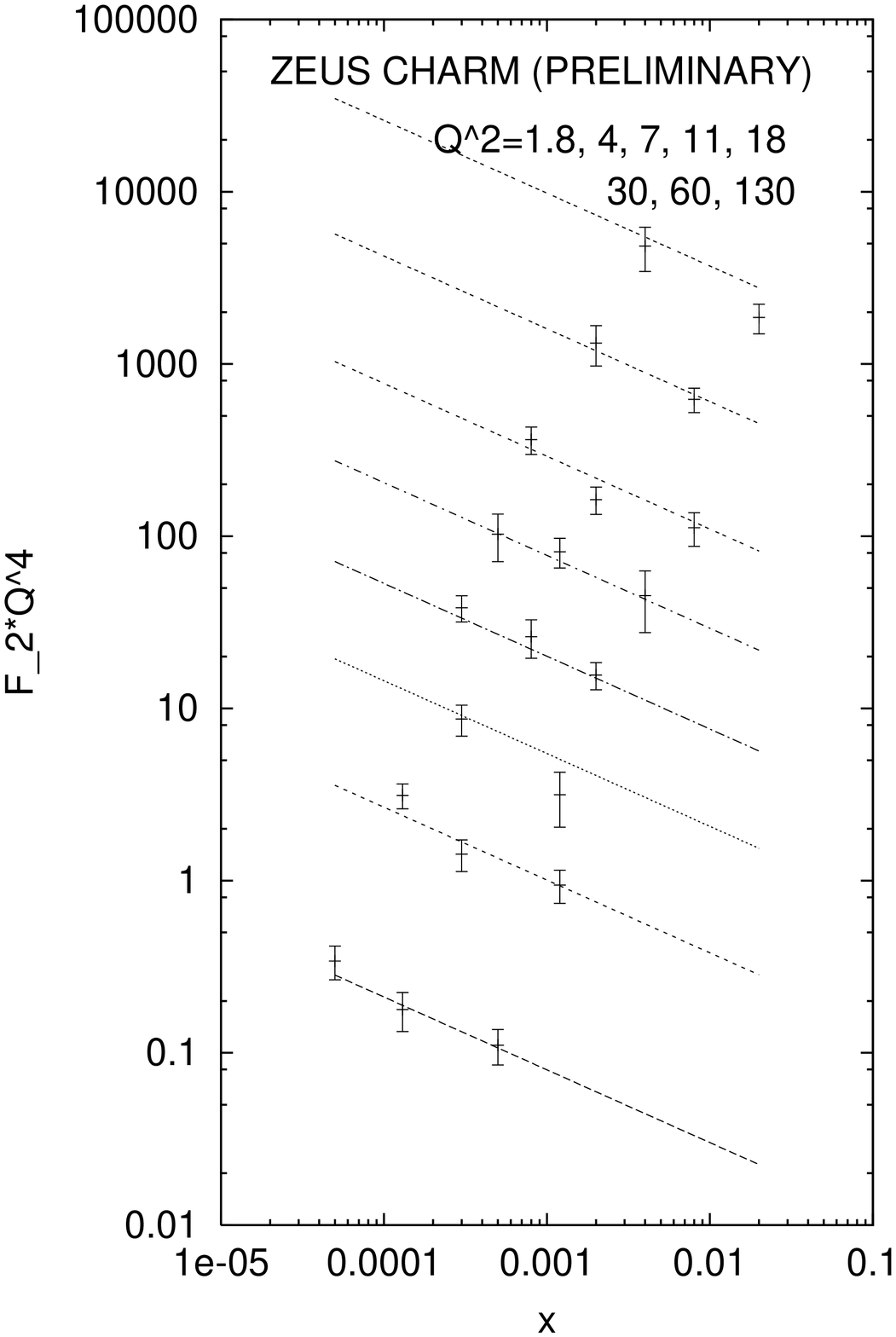}
\end{center}
\caption[]{Hard-pomeron fit to preliminary ZEUS data for $F_2^c$}
\begin{center}
\vskip 5mm
\includegraphics[width=.75\textwidth]{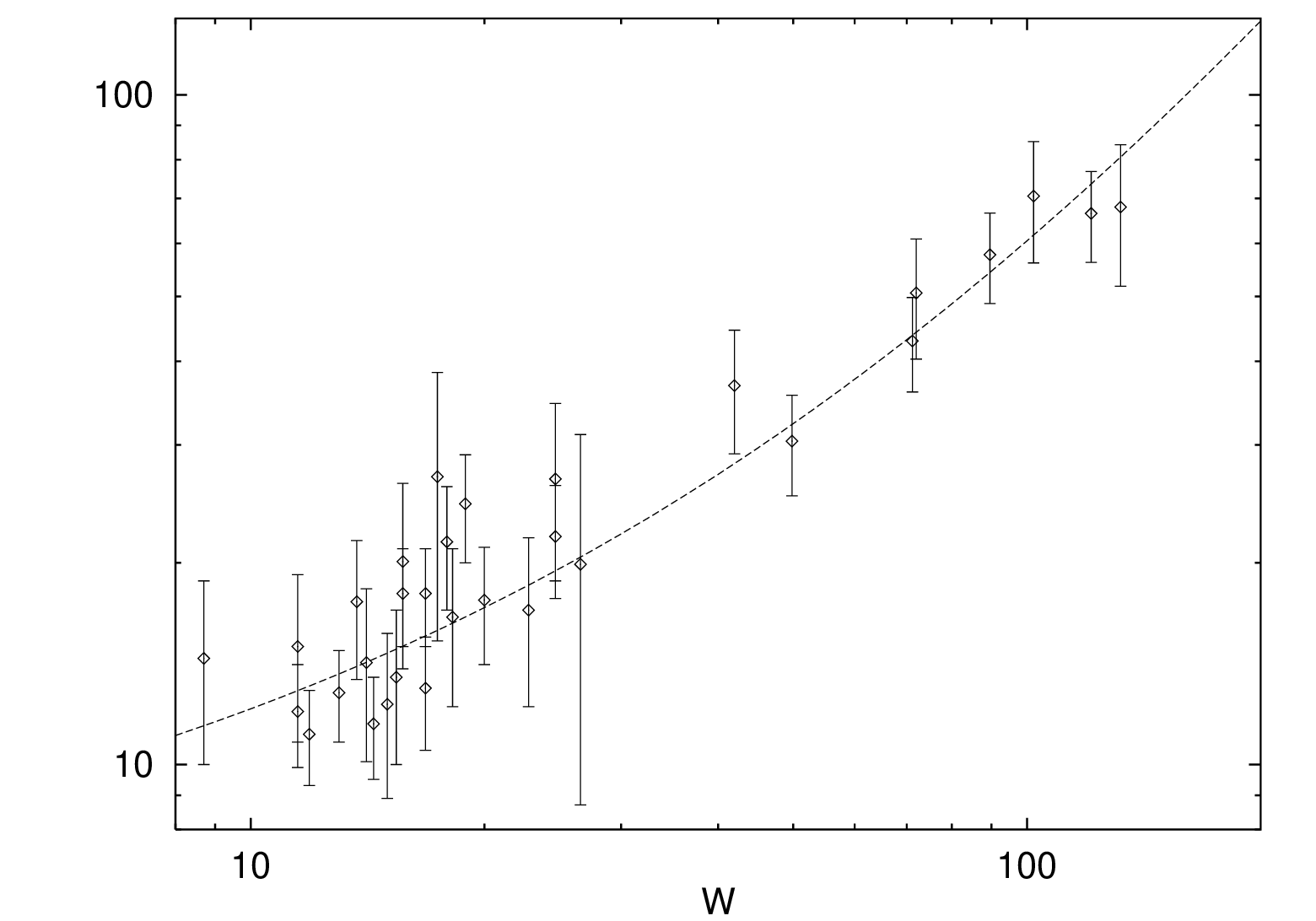}
\end{center}
\caption[]{Two-pomeron fit to data for $\gamma\,p\to J/\psi\,p$}
\label{jpsi}
\end{figure}

\section{Perturbative evolution}
At high $Q^2$, parton distributions evolve with $Q^2$ according to
a perturbative evolution equation, called the DGLAP equation.
In the singlet channel, the equation 
is written\cite{esw} in terms of a 2-component vector {\bf u} whose elements
are the singlet quark distribution and the gluon distribution:
\begin{equation}
\dot {\bf u}(x,t)=\int_x^1{dz\over z}\,{\bf P}(z,t){\bf u}(x/z,t)
\label{DGLAP}
\end{equation}
On the left-hand side, {\bf u} appears differentiated with respect to
\begin{equation}
t=\log (Q^2/\Lambda ^2)
\end{equation}
On the right-hand side {\bf P} is the $2\times 2$ splitting matrix.

The splitting matrix {\bf P}
may be expanded as power series in $\alpha_S(Q^2)$.
It was observed long ago\cite{yndurain} that if one uses just the first
term in this expansion, 
\begin{equation}
{\bf P}(z,t)={\beta_0\over \log(Q^2/\Lambda^2)}\,{\bf p}(z)
\label{approx}
\end{equation}
and assumes that each element of
${\bf u}(x,t)$ has simple
power behaviour in $x$ as in (\ref{f2}),
then the DGLAP equation gives a simple differential equation 
for the coefficient matrix ${\bf f}(Q^2)$ that multiplies the power. 
Its solution is
\begin{equation}
{\bf f}(Q^2)={\bf f}(\Lambda^2)\left(\log{Q^2\over\Lambda^2}\right)^
{\gamma(\e)}
\end{equation}
where $\gamma(N)$ is the eigenvalue of $\beta_0 {\bf p}(N)$.
Here ${\bf p}(N)$ is the  Mellin transform of ${\bf p}(N)$:
\begin{equation}
{\bf p}(N)=\int_0^1dz\,z^N\,{\bf p}(z)
\end{equation}

For $\e=0.4$  the eigenvalue ${\gamma(\e)}$ is close to 3. 
One may fit the data satisfactorily by requiring the hard-pomeron coefficient
function $f_0(Q^2)$ to have this behaviour at large $Q^2$, though not as well
as with the power behaviour of (\ref{coeff}). However, in any case it
is not valid to use the approximation (\ref{approx}) at low $x$,
even if it is supplemented by adding in higher-order terms in the
perturbative expansion. Such an expansion is not legal at small $x$,
even though it is widely used\cite{grv}.

The problem is that two of the elements $p_{qG}$ and
$p_{GG}$ of the matrix ${\bf p}(z)$ 
have poles\cite{esw} at $z=0$, so that their Mellin transforms have poles
at $N=0$. For example,
\begin{equation}
\pi\, p_{GG}(z)\sim {C_A\over z}~~~~~~~~~~~~~~~~\pi\, p_{GG}(N)\sim {C_A\over N}
\end{equation}
If one includes higher-order terms in the perturbative expansion,
all the elements of the Mellin transform of the splitting matrix are singular
at $N=0$. The consequence is that, as soon as $Q^2$ is large enough
for the DGLAP equation to be applicable, 
$Q^2>Q_0^2$ say, the Mellin transforms of the
parton distributions acquire rather nasty singularities at $N=0$, 
something like exp$(1/N)$.  This conflicts with what we believe we know about 
analyticity properties in $Q^2$. As I explained in section 2, the Regge
amplitude at high energy (or small $x$) is essentially just the Mellin 
transform: the relationship is
\begin{equation}
\ell =N+1
\end{equation}
So a singularity at $N=0$ corresponds to one at $\ell=1$. However, the
analyticity properties of the amplitude whose imaginary part is
$F_2(x,Q^2)$ were well studied nearly 40 years ago\cite{collins} and 
there was never any suggestion that it has
a singularity at $\ell=1$, let alone a nasty one.
If there is 
 no singularity when $Q^2$ is small, it is not possible that one suddenly
appears when we continue analytically in $Q^2$ to $Q^2>Q_0^2$.
That is, there can be no singularity of the splitting matrix at $N=0$ ---
and probably not, indeed, at any other value of $N$.

The appearance of the singularity at $N=0$ arises because of the
perturbation expansion. Compare, for illustration, the function
\begin{equation}
\phi(N,\alpha_S)=N-\sqrt{N^2-\alpha_S}
\end{equation}
Its expansion in powers of $\alpha_S$ is
\begin{equation}
\phi(N,\alpha_S)={\alpha_S\over 2N}+{\alpha_S^2\over 8N^3} +\dots
\label{phi}
\end{equation}
Each term in the expansion is singular at $N=0$, but manifestly the
complete function $\phi(N,\alpha_S)$ is not. The expansion is illegal
near $N=0$.

The timelike splitting matrix is very similar\cite{bass}
in form to $\phi(N,\alpha_S)$,
but in the spacelike region things are a little more complicated.
If one combines the DGLAP equation with the BFKL equation, which could
well be a valid thing to do when $x$ or $N$ is small, one finds
that the element $P_{GG}$ of the spacelike
splitting matrix is given\cite{jaros} in terms
of the Lipatov characteristic function\cite{bfkl} $\chi(\omega,\alpha_S)$:
\begin{equation}
\chi(P_{gg}(N,t),\alpha_S)=N
\label{P}
\end{equation}
To lowest order in $\alpha_S$, 
\begin{equation}
\chi(\omega,\alpha_S)={3\alpha_s\over\pi}\left [2\psi(1)-\psi(\omega)-
\psi(1-\omega)\right ]
\label{chi}
\end{equation}
One may easily verify\cite{cudell} that, if one inserts (\ref{chi}) into
(\ref{P}), each term in the expansion of $P_{gg}(N,t)$ in powers of
$\alpha_S$ is singular at $N=0$, but the complete function $P_{gg}(N,t)$
is finite there. Some authors\cite{kellis} therefore advocate that one should
use the solution to the full equation (\ref{P}) and avoid expanding it.

This resummation
is obviously sensible, but there are still serious unsolved problems.
Firstly, it is highly doubtful that it is valid to use the uncorrected BFKL
equation\cite{cl}: it assumes that nonperturbative effects cause
no complications, and does not properly take account of energy conservation.
Secondly, even if one ignores these difficulties, using the lowest-order
approximation (\ref{chi}) to $\chi(\omega,\alpha_S)$ is not valid: the
next-to-leading-order correction is huge\cite{fl}. There have recently
been some interesting attempts to solve this problem\cite{ciaf}, but
more work remains to be done. 

So at present, although we know that the elements of the splitting matrix
cannot diverge at $N=0$, we cannot calculate them. Any application of the
DGLAP equation in which they are expanded perturbatively, and in which
use is made of the fact that the terms in this expansion are large,
cannot be trusted.

\section{Summary}
\begin{itemize}
\item When we use the DGLAP equation at small $x$ it is not valid to
use an unresummed perturbative expansion of the splitting matrix ---
at least, not until $Q^2$ is somewhat larger than is normally assumed. At
present, we do not know how to perform the necessary resummation
properly.
\item Regge theory with two pomerons fits the small-$x$ data for 
$F_2(x,Q^2)$ extremely well up to the highest available values of $Q^2$.
It strongly suggests that
at, say, $Q^2=5$ GeV$^2$, most of $F_2(x,Q^2)$ at small $x$ 
is higher twist.
\item Perturbative QCD and Regge theory are not rival theories. They
complement each other and we have to learn how to make them fit together.
\end{itemize}

\font\eightit=cmti8
\noindent{\eightit
This research is supported in part by the EU Programme
``Training and Mobility of Researchers", Networks
``Hadronic Physics with High Energy Electromagnetic Probes"
(contract FMRX-CT96-0008) and
``Quantum Chromodynamics and the Deep Structure of
Elementary Particles'' (contract FMRX-CT98-0194),
and by PPARC}


\begin{thebibliography}{16}
%
\addcontentsline{toc}{section}{References}


\bibitem{collins} P D B Collins, {\it Introduction to Regge Theory},
Cambridge University Press (1977)
\bibitem{diffdis} A Donnachie and P V Landshoff,
Nuclear Physics B244 (1984)  322
\bibitem{sigtot} A Donnachie and P V Landshoff,
Physics Letters B296 (1992) 227
\bibitem{twopom}A Donnachie and P V Landshoff, Physics Letters B437 (1998) 408
\bibitem{bfkl}E A Kuraev, L N Lipatov and V Fadin, Soviet Physics JETP 45 (1977)
\bibitem{cl}J C Collins and P V Landshoff, Physics Letters B276 (1992) 196;
M F  McDermott, J R  Forshaw and G G  Ross,  Physics Letters B349 (1995) 189
\bibitem{fl}V S Fadin and L N Lipatov, Physics Letters B429 (1998) 127;
G Camici and M Ciafaloni, Physics Letters B430 (1998) 349
\bibitem{esw}R K Ellis, W J Stirling and B R Webber, 
{\it QCD and Collider Physics}
Cambridge University Press (1996)
\bibitem{cudell} J R Cudell, A Donnachie and P V Landshoff,
Physics Letters B448 (1999) 281
\bibitem{ballforte} R D Ball and S Forte, Physics Letters B405 (1997) 317
\bibitem{grv}M Gl\"uck, E Reya and  A Vogt, 
European Physics J C5 (1998) 461;
A D  Martin, R G  Roberts, W J  Stirling and R S Thorne, 
European Physics J C4 (1998) 463; CTEQ collaboration -- H.L. La et al,
hep-ph/9903282 
\bibitem{elop}R J Eden, P V Landshoff, D I Olive and J C Polkinghorne,
{\it The Analytic S-Matrix},
Cambridge University Press (1966)
\bibitem{capella}A  Capella, A  Kaidalov, C  Merino, D  Pertermann, 
and J  Tran Thanh Van, European Physics J C5 (1998) 111;
E Gotsman, E Levin, U Maor and E Naftali, hep-ph/9904277;
K Golec-Biernat and  M W\"usthoff, hep-ph/9903358 
\bibitem{ermolaev} B Ermolaev, private communication
\bibitem{levin}E M Levin, Nuclear Physics B453 (1995) 303
\bibitem{sotiropoulos}K D Anderson, D A Ross and  M G Sotiropoulos,
Nuclear Physics B515 (1998) 249
\bibitem{yndurain}C Lopez and F J Yndurain, Nuclear Physics B183 (1981) 157
\bibitem{bass}A Bassetto, M Ciafaloni and G Marchesini, 
Nuclear Physics B(1980)477
\bibitem{jaros}T Jaroszewicz, Physics Letters 116B (1982) 291;
 S Catani and F Hautmann, Nuclear Physics B427 (1994) 475
\bibitem{kellis}R K  Ellis, F  Hautmann and B R  Webber,
Physics Letters B348 (1995) 582; R S Thorne, hep-ph/9901331
\bibitem{ciaf}M Ciafaloni, D Colferai and  G P Salam,
 hep-ph/9905566; R D Ball and S Forte, hep-ph/9906222  

\end{thebibliography}
\end{document}